# A Simulation Methodology for Superconducting Qubit Readout Fidelity

Hiu Yung Wong[*], Yaniv Jacob Rosen[+], Kristin M. Beck[+], and Prabjot Dhillon
Electrical Engineering, San Jose State University, CA, USA
[+]Lawrence Livermore National Laboratory, Livermore, CA, USA
[*]hiuyung.wong@sjsu.edu

**ABSTRACT**

**Qubit readout is a critical part of any quantum computer including the superconducting-qubit-based one. The readout fidelity is affected by the readout pulse width, readout pulse energy, resonator design, qubit design, qubit-resonator coupling, and the noise generated along the readout path. It is thus important to model and predict the fidelity based on various design parameters along the readout path. In this work, a simulation methodology for superconducting qubit readout fidelity is proposed and implemented using Matlab and Ansys HFSS to allow the co-optimization in the readout path. As an example, parameters are taken from an actual superconducting-qubit-based quantum computer and the simulation is calibrated to one experimental point. It is then used to predict the readout error of the system as a function of readout pulse width and power and the results match the experiment well. It is found that the system can still maintain high fidelity even if the input power is reduced by 7dB or if the readout pulse width is 40% narrower. This can be used to guide the design and optimization of a superconducting qubit readout system.**

## I. INTRODUCTION

Superconducting qubits are one of the most promising quantum computing architectures [1]. While a qubit needs to have enough isolation to achieve a long coherence time, it should also be allowed to interact with the outside world for the readout operation. Often, a resonator is coupled to a qubit to allow dispersive readout, in which the resonator will experience a resonant frequency shift depending on the final state of the qubit [2]. This frequency shift is called the Cross-Kerr, $\chi$. The larger the $\chi$, the easier it is to distinguish the qubit's $|0\rangle$ and $|1\rangle$ states. However, this will also result in a shorter coherence time. The distinguishability of the $|0\rangle$ and $|1\rangle$ states also depends on the readout pulse power and duration, the resonator scattering matrix, and the noise from the circuits. Therefore, it is important to co-optimize the resonator design, qubit-resonator coupling, and reading pulse length and power with the noise taken into account.

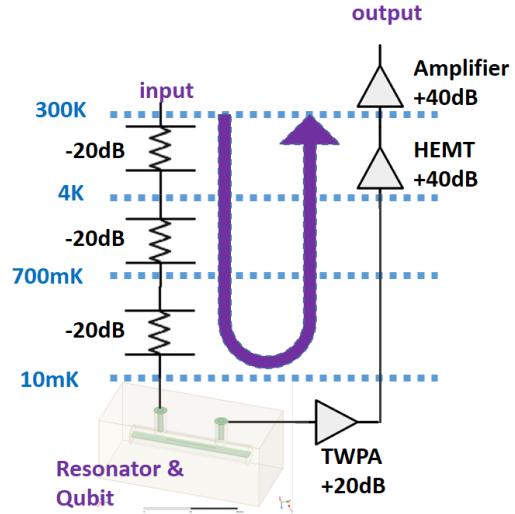

Figure 1: The qubit system used. The readout path is highlighted.

In this paper, a simulation framework and methodology are proposed and implemented using Matlab and Ansys HFSS. Calibration is done using one experimental data point and it can be used to predict how the fidelity changes with the readout pulse width and power.

## II. THE QUBIT READOUT SYSTEM

Fig. 1 shows the experimental hardware system used in this paper. Quantum Machine OPX is used as the control hardware, with a single sideband mixer and stable RF source used to upconvert the outputs to the qubit and readout frequencies [3]. Nominal power of -47dBm 3.5μs ($t_p$) readout pulse of 7.246245GHz is used. After three attenuation stages (-60dB in total) and the attenuation due to the cables (measured to be -16dB), the pulse reaches the input port (port 1, where the pulse becomes -123dBm) of the resonator coupled to a qubit at 10mK. The qubit is tantalum-based with a high coherence time (~0.25ms) [4]. The signal from the output port (port 2) of the resonator is then amplified by a Traveling Wave Parametric Amplifier (TWPA) (+20dB) at 10mK, a High Electron Mobility Transistor (HEMT) amplifier at 4K (+40dB), and a 300K amplifier (+40dB). Quadrature measurement is performed on the amplified output signal, which represents the $S_{21}$ of the resonator/qubit system, to distinguish the qubit |0⟩ and |1⟩ states. The χ of the system is measured to be 114kHz.

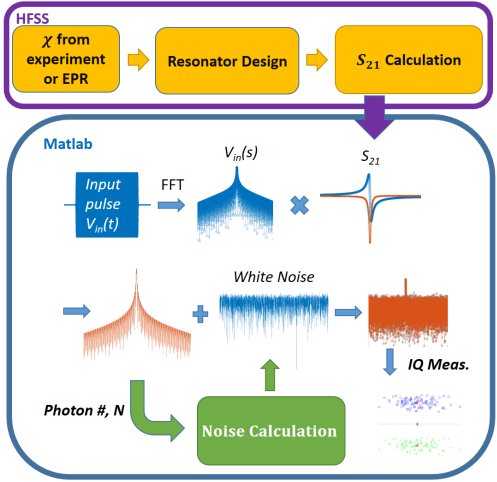

Figure 2: Ilustration of the simulation flow.

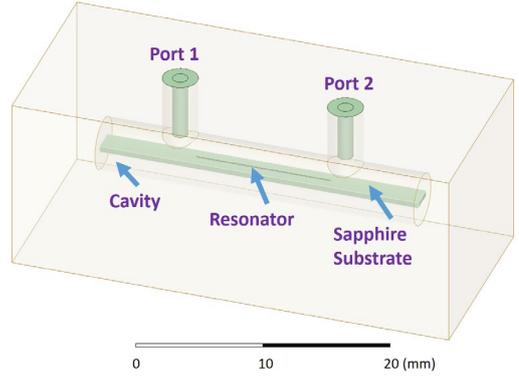

Figure 3: The cavity and resonator used in the HFSS simulation.

III. SIMULATION METHODOLOGY

Fig. 2 shows the simulation framework. The framework uses Ansys HFSS [5] to perform the scattering matrix simulation of the resonator (to be detailed in the next section). The $S_{21}$ obtained is then fed into a MATLAB program to simulate the readout process. There are three major noise sources. The first one is the quantum noise due to the photon number fluctuation after the resonator. The second one is the noise due to the TWPA. Since TWPA is a quantum-limited amplifier, therefore, at the best case, it only reduces the signal-to-noise ratio by half when the input is a single photon [6]. This is equivalent to adding 3dB of noise to its output. Thirdly, the two low noise amplifiers contribute thermal noise equivalent to $T_{eff}$ = 1.5K [7] and $T_{eff}$ = 54K [8], respectively, with a noise spectral density of $4kT_{eff}R$, where $k$ is the Boltzmann constant and $R$ is 50Ω.

In [9], qubit readout quantum noise (relative to the distance between the |0⟩ and |1⟩ states) was derived based on the qubit relaxation time, resonator photon lifetime, quantum-limited amplifier noise effective temperature, etc. However, this does not allow the inclusion of other noise sources.

To allow the simulation of the quantum noise in our classical framework, the quantum noise due to the photon fluctuation and coming from the TWPA are modeled with white noise, and the fundamental quantum noise limit of a linear amplifier is used based on [10]. The associated equivalent noise temperature, $T_n$, is computed using the following equation derived in [10],

$$T_n = \frac{1}{\ln 2} \frac{hf}{k} \quad (1)$$

where $h$, $f$, and $k$, are the Planck's constant, pulse frequency, and Boltzmann's constant, respectively. A white noise corresponding to $T_n$ is used in the simulation. $T_n$ is found to be 0.5K.

The white noise power spectral density has a unit of $dBm/Hz$. It is converted to power in $dBm$ by multiplying by $B/t_p$, where $B$ is a unitless fitting parameter. Therefore, the noise power is $kT_nRB/t_p$.

All noises are generated in the time domain and converted into the frequency domain using Fast Fourier Transformation to be added to the signal.

The output pulse from the resonator is simulated by multiplying the attenuated input pulse and the $S_{21}$ of the resonator in the frequency domain. The total noise is then added to the output pulse. The real and imaginary parts at the readout frequency are extracted to simulate the quadrature measurement. 1000 random runs are performed to obtain the statistics.

IV. SIMULATION SETUP

Since the experimental χ is available, the resonators are designed to have eigenfrequencies of 7.252456GHz and 7.252612GHz, to emulate the coupled qubit's |0⟩ and |1⟩ states, respectively. This is achieved by designing a resonator length of 3.29265mm and 3.2925mm, respectively, without simulating the qubit. Dense mesh is required to achieve the required accuracy. This gives an effective χ of 156kHz, which is similar to that of the hardware. Fig. 3 shows the design of the cavity and the resonator with Q ~ 48k, similar to the experimental value. If experimental χ is not available, it can be obtained using the Energy Participation Ratio (EPR) method with HFSS [2] for the qubit design and device layout. The readout pulse frequency is $f$ = 7.252534 GHz, which is the average of the two resonator frequencies. Based on the simulation, the number of photons entering port 1 is about 363 and 94 photons are emitted from port 2.

Fig. 4 shows the output signal before and after the chain of amplifiers for resonator coupled with qubit with states |0⟩ and 1⟩. It can be seen that the noise reduces the distinguishability.

## V. SIMULATION RESULTS

At $t_p = 3.5\mu s$, to obtain the best matching between the experimental and the simulation results, B=3500 is used. Fig. 5 shows the fidelity of the qubit readout based on experimental quadrature measurement and simulation with this offset. I-Q distributions are plotted for the two qubit states (|0> and |1>) for 1000 samples and each blob represents the spreading of the I-Q signal when the qubit is at |0⟩ or |1⟩ state, respectively. It shows that the simulation and experimental results match each other pretty well in terms of |0⟩/|1⟩ blob center distance to blob spreading ratio. Note that in the experiment, there are some errors that do not follow the Gaussian distribution (e.g. green cross inside the blue |0⟩ blob). They are believed to be qubit reset errors that are dependent on the measurement fidelity and are not captured in the simulation. Before every measurement, the qubit needs to be set up at the correct state. This is done by measuring the qubit first and then applying a setup pulse, if needed, to rotate the qubit to the required state. If this is not done properly, there will be qubit reset errors. In the simulation, this is not simulated.

This calibrated framework is then used to study how the input pulse power and pulse width change the fidelity of qubit readout. Fig. 6 shows that when the pulse width is reduced to 2μs, the |0⟩/|1⟩ I-Q distributions merge and errors are expected to increase substantially. Fig. 7 shows the error as a function of the readout pulse width. When the pulse width is less than 2μs, which is 40% of the nominal pulse width, the error increases. Note that the experiment has non-zero errors for long pulse widths and this is due to the reset error as mentioned earlier. Also, the experiment error increases faster than the simulation one when the pulse width reduces. This is due to the fact that the corresponding shorter pulses are used to read the qubit before resetting during the experiment. Shorter pulses have larger readout error and thus causes more reset errors. The purpose of this simulation is not to match the error quantitatively but to predict when the error will increase substantially. This is because once the error starts increasing when the I-Q blobs merge, the qubit is not suitable for fault-tolerant computation anymore. Therefore, predicting when the I-Q blob merge is the primary goal.

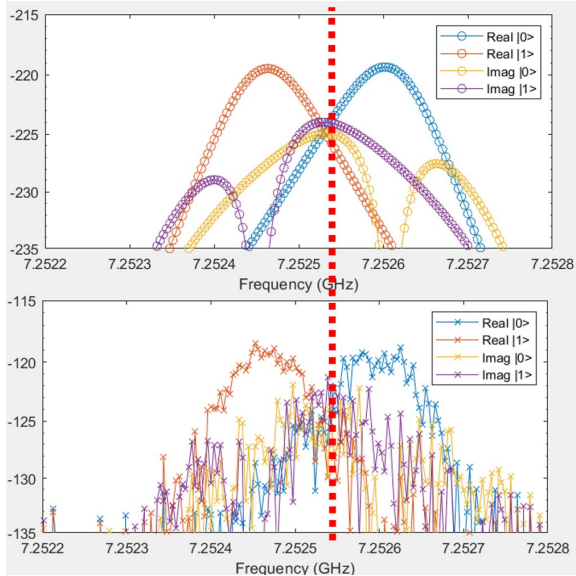

Figure 4: The real and imaginary compoents of the signal after the resonantor before adding the quantum noise (Top) and after the amplification chain in Fig. 1 (Bottom). The red dotted line indicates the reading pulse frequency.

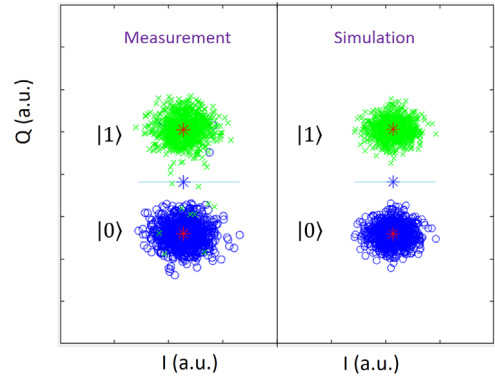

Figure 5: The quadrature measurement (Left) and simulation (Right) for reading |0⟩ and 1⟩ states, with readout pulse at nominal power and $t_p = 3.5\mu s$

The same setup is then used to predict how the readout pulse power affects the readout error. Fig. 8 shows that the I-Q blobs merge in both simulation and experiment when the power is 8dB less than the nominal power, at which the error is expected to increase substantially. Fig. 9 shows the experiment and simulation errors as a function of the relative readout pulse power (relative to the nominal power). As expected, both simulation and experiment predict the errors increase substantially after -7dB. Similar to the previous case, the experiment has non-zero errors at large pulse power due to reset error. The reset error is exacerbated when the I-Q blobs merge because lower power has larger readout errors and thus induces more reset errors.

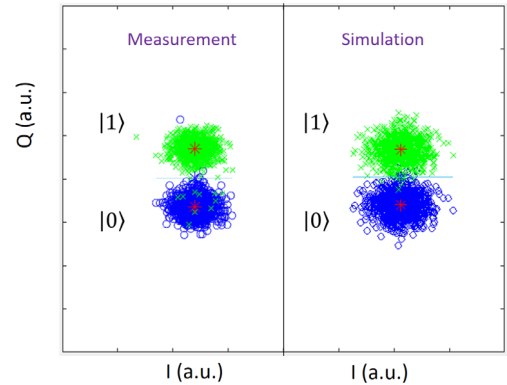

Figure 6: The quadrature measurement (Left) and simulation (Right) for reading |0⟩ and 1⟩ states, with nominal readout power and $t_p = 2.0\mu s$

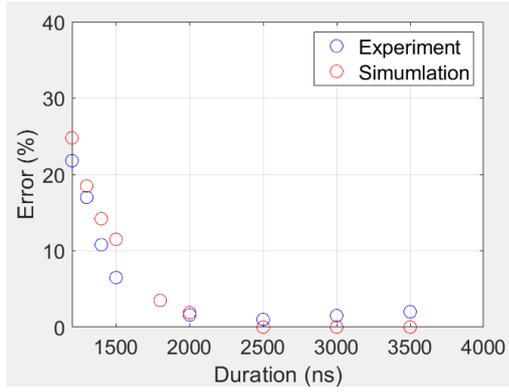

Figure 7: Simulated and meaasured readout errors of the qubit readout system as a function of readout pulse duration.

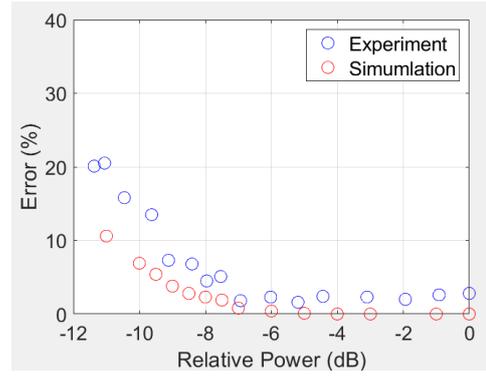

Figure 9: Simulated and measured readout errors of the qubit readout system as a function of readout pulse power relative to the nominal power.

## VI. CONCLUSIONS

In this paper, a simulation method for predicting superconducting qubit readout fidelity is proposed and implemented using Matlab and HFSS. The model is first calibrated to one experimental data point and then it is used to predict how the readout pulse width and pulse power change the readout error. After the calibration, the simulation results match the experi-mental result well and can predict when the error will increase substantially. It is found that the pulse width can be reduced by 50% or the pulse power can be reduced by 7dB while maintaining high fidelity for the system being studied. The system can thus be further optimized accordingly.

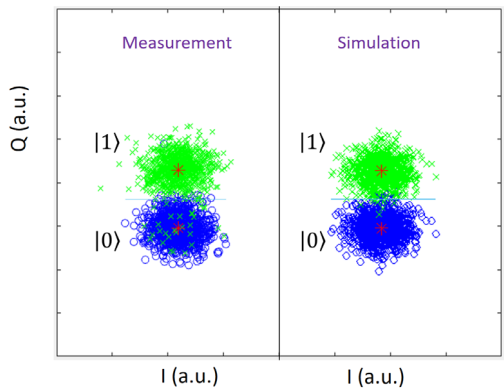

Figure 8: The quadrature measurement (Left) and simulation (Right) for reading $|0\rangle$ and $1\rangle$ states, with a readout pulse power 8dB less than the nominal power and $t_p = 3.5\mu s$

## VII. ACKNOWLEDGMENT

This material is based upon work supported by the National Science Foundation under Grant No. 2125906. The authors thank MITLL and IARPA for allowing us to use their TWPAs. Prepared in part by LLNL under Contract DE-AC52-07NA27344.